\documentclass[twoside,12pt]{article}  %%% 这是 LaTeX 格式文件必须的指令，不可缺少
\usepackage{CJK}   %%%%% 使用中文、日文、韩文时调用此宏包
\usepackage{indentfirst}  %% 节标题后第一行段缩进
\usepackage{bm}    %%%%%  排印数学符号黑斜体时调用此宏包
\usepackage{graphicx}  %%%% 文中插入 eps 图形时调用此宏包，插图方法 1
\begin{document}    %% 文本文件开始，这是必须的指令

\begin{CJK*}{GBK}{song}  %% 开始进入中文环境

%-------------------  First Head  -----------------------------------------
\thispagestyle{empty} \vspace*{0.8cm}\hbox
to\textwidth{\vbox{\hfill\huge\sf \hfill}}
\par\noindent\rule[3mm]{\textwidth}{0.2pt}\hspace*{-\textwidth}\noindent
\rule[2.5mm]{\textwidth}{0.2pt}

%=================== Text begin here =============================================

\begin{center}
\LARGE\bf One dimensional lazy quantum walks and occupancy rate$^{*}$   %% 论文题目
\end{center}

\footnotetext{\hspace*{-.45cm}\footnotesize $^\dag$Corresponding author. E-mail: lidansusu007@163.com }

\begin{center}
\rm Dan Li $^{\rm a,b)\dagger}$, \ \ Michael Mc Gettrick$^{\rm b)}$, \ \ Wei-Wei Zhang $^{\rm a)}$, \ and  \ Ke-Jia Zhang $^{\rm a)}$
\end{center}

\begin{center}
\begin{footnotesize} \sl
${State\ Key\ Laboratory\ of\ Networking\ and\ Switching\ Technology}$, \\${Beijing\ University\ of\ Posts\ and\ Telecommunications,\ Beijing,\ 100876,\ China}^{\rm a)}$ \\   %%%% 地址 a)
${The\ De\ Brun\ Centre\ for\ Computational\ Algebra,\ School\ of\ Mathematics, Statistics\ and\ Applied\ Mathematics}$, \\${National\ University\ of\ Ireland,\ Galway}^{\rm b)}$ \\   %%%% 地址 b)

%%% 更多地址依次往下延续
\end{footnotesize}
\end{center}

\begin{center}
\footnotesize (Received X XX XXXX; revised manuscript received X XX XXXX)
          %% (Received 日 月 年; revised manuscript received 日 月 年)
\end{center}

\vspace*{2mm}

\begin{center}
\begin{minipage}{15.5cm}
\parindent 20pt\footnotesize
Lazy quantum walks were presented by Andrew M. Childs to prove that the continuous-time quantum walk is a limit of the discrete-time quantum walk [Commun.Math.Phys.294,581-603(2010)]. In this paper, we discuss properties of lazy quantum walks. Our analysis shows that lazy quantum walks have $O(t^n)$ order of the n-th moment of the corresponding probability distribution, which is the same as that for normal quantum walks. Also, the lazy quantum walk with DFT (Discrete Fourier Transform) coin operator has a similar probability distribution concentrated interval to that of the normal Hadamard quantum walk. Most importantly, we introduce the concepts of occupancy number and occupancy rate to measure the extent to which the walk has a (relatively) high probability at every position in its range. We conclude that lazy quantum walks have a higher occupancy rate than other walks such as normal quantum walks, classical walks and lazy classical walks.
%%%% 论文摘要
\end{minipage}
\end{center}

\begin{center}
\begin{minipage}{15.5cm}
\begin{minipage}[t]{2.3cm}{\bf Keywords:}\end{minipage}
\begin{minipage}[t]{13.1cm}
%%%%% 关键词
Lazy quantum walk,  Occupancy number, Occupancy rate
\end{minipage}\par\vglue8pt
{\bf PACS: 03.67.Ac, 03.67.Lx, 02.30.Nw}
%%% PACS 分类码
%% 查询网址：http://www.aip.org/pacs
\end{minipage}
\end{center}

\section{Introduction}
\label{sec:level1}

Due to constructive quantum interference along the paths in the discrete or the continuous version, quantum walks \cite{000,001} provide a method to explore all possible paths in a parallel way. Recently, algorithms based on quantum walks have been established as a dominant technique in quantum computation. Up to now, algorithms based on quantum walks have a large number of applications, ranging from element distinctness \cite{110} to database searching \cite{111,112,113,114}, from constructing quantum Hash schemes \cite{115,116} to graph isomorphism testing \cite{117,118}.

There are many kinds of quantum walk models, such as single-particle quantum walks \cite{120,121,128,503}, two-particle quantum walks \cite{122,123,124}, three-state quantum walks \cite{002}, controlled interacting quantum walks  \cite{115,116}, indistinguishable particle quantum walks \cite{125,126}, disordered quantum walk \cite{129,127} etc. Each type of quantum walk has its own special features and advantages.

Lazy quantum walks were presented by Andrew M. Childs to prove that the continuous-time quantum walk is a limit of the discrete-time quantum walk \cite{130}. Childs constructs a lazy quantum walk, which only takes a step with a small probability, to obtain small eigenvalues. So the discrete-time quantum walk whose behavior reproduces that of the continuous-time quantum walk is the appropriate limit.  Also much attention  has been attached in one-dimensional three-state quantum walks recently \cite{002,301,302,304,305,306}. They present general three-state quantum walks and discuss limit distribution of these walks.  In this paper, we will consider general properties of discrete lazy quantum walks, i.e. three-state quantum walks,  such as the order of the moments of the probability distribution, the probability distribution concentrated interval and the entanglement between the coin and position of the particle.

Quantum walks have special properties, such as ballistic evolution and high probability of reaching remote points. This is because quantum walks create a superposition of all potential routes, furthermore creating coherent states over these routes. For Hadamard quantum walks on the line, the probability distribution concentrated interval is $[-(\frac{1}{\sqrt{2}}+\varepsilon)t,(\frac{1}{\sqrt{2}}+\varepsilon)t]$ \cite{120,121}. However, previous work has only considered the overall probability distribution concentrated interval, not the individual probabilities at the positions in the interval. Therefore, we present the occupancy number and occupancy rate as a measure of the particle's  occupancy.

The paper is structured as follows. In Sect.\ref{sec:level2} , we introduce the specific mathematical formalism for lazy quantum walks on the line. In Sect.\ref{sec:level3} , we study the probability distribution concentrated interval and the moments of the probability distribution for lazy quantum walks, and prove that these moments have the same orders as those for normal quantum walks. We define and analyze the occupancy number and occupancy rate in  Sect.\ref{sec:level4}  and we discuss the entanglement between the position and coin for lazy quantum walks in Sect.\ref{sec:level5}. Finally, a short conclusion is given in Sect.\ref{sec:level6}.

\section{Lazy Quantum Walks }
\label{sec:level2}

 In general, quantum walks on the line have two directions to move, right and left. But lazy quantum walks have three choices, right, left and stay put. This results in different behaviour compared to their non-lazy counterparts. In this section, we introduce the specific mathematical formalism for lazy quantum walks on the line.

Lazy quantum walks take place in the product space $\mathcal{H}={\mathcal{H}}_{p}\bigotimes {\mathcal{H}}_{c}$. ${\mathcal{H}}_{p}$ is a Hilbert space which has orthonormal basis given by the position states $\{|x\rangle, x\in\mathcal{Z}\}$. The default initial position state is $|0\rangle$. Due to the three choices of the movement, lazy quantum walks have a three-dimensional coin. Therefore, $\mathcal{H}_{c}$ is a Hilbert space spanned by the orthonormal basis $\{|r\rangle,|s\rangle,|l\rangle\}$ ($r$ for right, $s$ for stay put, $l$ for left).

Let $|x,\alpha\rangle$ be a basis state, where $x\in \mathcal{Z}$ represents the position of the particle and $\alpha\in \{r,s,l\}$ represents the coin state. The evolution of the whole system at each step of the walk can be described by the global unitary operator, denoted by $U$,
\begin{equation}\label{21}
    U=S({\mathcal{I}}\otimes C),
\end{equation}where $S$ is defined by
\begin{equation}\label{22}
    S=|x+1,r\rangle\langle x,r|+|x,s\rangle\langle x,s|+|x-1,l\rangle\langle x,l|.
\end{equation}$I$ is the identity matrix which operates in $\mathcal{H}_{p}$, while $C$ is the coin operation. In this paper, we consider two kinds of coin operators. The first kind is the DFT (Discrete Fourier Transform) coin operator
\begin{equation}\label{23}
    C=\frac{1}{\sqrt{3}}\left(
          \begin{array}{ccc}
            1 & 1 & 1 \\
            1 & e^{\frac{2\pi i}{3}} & e^{\frac{4\pi i}{3}} \\
            1 & e^{\frac{4\pi i}{3}} & e^{\frac{2\pi i}{3}} \\
          \end{array}
        \right).
\end{equation}Besides this coin operator, there are other kinds of $3\times3$ coin operators \ref{301,304}
\begin{equation}\label{24}
    G(\rho)=\left(
          \begin{array}{ccc}
            -\rho^2 & \rho \sqrt{2-2\rho^2} & 1-\rho^2 \\
            \rho \sqrt{2-2\rho^2} & 2 \rho^2-1 & \rho \sqrt{2-2\rho^2} \\
            1-\rho^2 & \rho \sqrt{2-2\rho^2} & -\rho^2 \\
          \end{array}
        \right)
\end{equation}
with the coin parameter $\rho\in (0,1)$. This coin operator is equal to the Grover operator when $\rho=\sqrt{\frac{1}{3}}$. We will call the parameter $\rho$ the \emph{laziness parameter} of the lazy quantum walk. Since the position after applying the shifting operator $S$ depends on the coin state of the particle, the walk will generate coin-position entanglement \cite{201}.

\section{Large $t$ Behavior of Lazy Quantum Walks }
\label{sec:level3}

Fourier analysis is a powerful tool and the right tool to exploit the symmetry of quantum walks.  To compare lazy quantum walks with normal quantum walks, in this section, we use Fourier analysis to analytically study the large $t$ behavior of lazy quantum walks.

Firstly, we define the state of the walker, $\Psi$, at time $t\in \mathcal{N}$ and position $x\in \mathcal{Z}$ to be a 3-dimensional vector. We denote this as
\begin{equation}\label{31}
    \Psi(x,t)=\left(
                  \begin{array}{c}
                    \psi_0(x,t) \\
                    \psi_1(x,t) \\
                    \psi_2(x,t) \\
                  \end{array}
                \right).
\end{equation}

Let
\begin{equation}\label{32}
    \widehat{\Psi}(k,t)=\sum_x e^{ikx} \Psi(x,t),
\end{equation}with the inverse transform given by
\begin{equation}\label{33}
    \Psi(x,t)=\int_{-\pi}^{\pi} \frac{dk}{2\pi} \widehat{\Psi}(k,t) e^{-ikx}.
\end{equation}

In momentum space, i.e. the space of $\widehat{\Psi}(k,t)$, the shift operator becomes
\begin{equation}\label{34}
    \widehat{S}=\left(
      \begin{array}{ccc}
        e^{ik} & 0 & 0 \\
        0 & 1 & 0 \\
        0 & 0 & e^{-ik} \\
      \end{array}
    \right).
\end{equation} So the walker evolves by the relation
\begin{equation}\label{35}
    \widehat{\Psi}(k,t+1)= \widehat{U} \widehat{\Psi}(k,t),
\end{equation}where $\widehat{U}=\widehat{S}\cdot C$. Therefore,
\begin{equation}\label{36}
    \widehat{\Psi}(k,t)= \widehat{U}^t \widehat{\Psi}(k,0),
\end{equation} It is worth mentioning that $ \widehat{\Psi}(k,0)=\Psi(0,0)$. For simplicity, we write $ \widehat{\Psi}(k,0)$ as  $\widehat{\Psi}_0$.

As a unitary matrix, $\widehat{U}$ has three eigenvalues. If $\widehat{U}$ possess apart from continuous spectrum also an isolated eigenvalue, i.e.  the eigenvalue is independent of $k$, the walker will have non-vanishing probability to stay at any position even in the limit of infinite number of steps \cite{002,301,302,304,305,306}.

Next, we will study the large $t$ behavior of lazy quantum walks, specifically the probability distribution concentrated interval and the moments of the probability distribution of the lazy quantum walk.

\begin{figure}[hbt]
  \center
  \includegraphics[width=6cm]{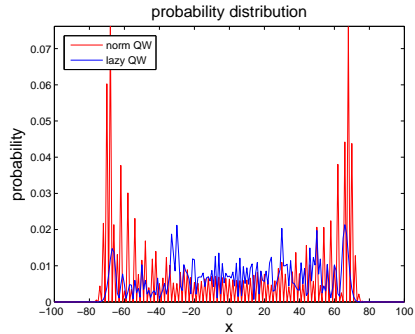}\\
  \center
  \caption{ Probability distribution of the lazy quantum walk and the normal quantum walk. The initial states are $[\sqrt{0.85};0;-\sqrt{0.15}]$ and $[\sqrt{0.85};-\sqrt{0.15}]$ respectively. }\label{F31}
\end{figure}

In Fig. \ref{F31}, we firstly show the probability distributions of the lazy quantum walk and the normal quantum walk. We chose the initial states $[\sqrt{0.85};0;-\sqrt{0.15}]$ / $[\sqrt{0.85};-\sqrt{0.15}]$ for lazy / normal quantum walks respectively, because they can produce symmetrical probability distributions. The choice of the initial state is important in studies of quantum walks, because interference features sensitively depend on the choice of the coin state. On the other hand, general properties do not depend on the choice of the initial coin state, so the coin state chosen here is not critical. The coin operators are DFT matrix and hadamard matrix respectively. We should remind the reader that the $2\times2$ DFT coin operator is precisely the hadamard matrix, which is the only quantum walk with an unbiased coin operator \cite{003}. For comparative purposes, we prefer to use the DFT coin operator in this paper.

From Fig. \ref{F31}, we can see that even though the two kinds of quantum walks have different probability distributions, the probability distribution concentrated intervals for the two quantum walks are close to each other. In papers \cite{120,121}, by using the method of stationary phase, the authors reported the conclusion that for hadamard quantum walks, the wave function is almost completely contained in the interval $[(-\frac{1}{\sqrt{2}}-\epsilon)t,(\frac{1}{\sqrt{2}}+\epsilon)t]$ and shrinks quickly outside this region. In papers \cite{301}, we can also know that the probability distribution concentrated interval of quantum walks with coin operator $G(\rho)$ is  $[(-\rho+\epsilon)t,(\rho+\epsilon)t]$. Specially, the lazy quantum walk with Grover coin operator has probability distribution concentrated interval $[(-\frac{1}{\sqrt{3}}+\epsilon)t,(\frac{1}{\sqrt{3}}+\epsilon)t]$ \cite{002}. Summarizing, we conclude that the unbiased normal quantum walk and  lazy quantum walks we consider in this paper have the same order of probability distribution concentrated intervals.

\begin{figure}[hbt]
    \center
  \includegraphics[width=6cm]{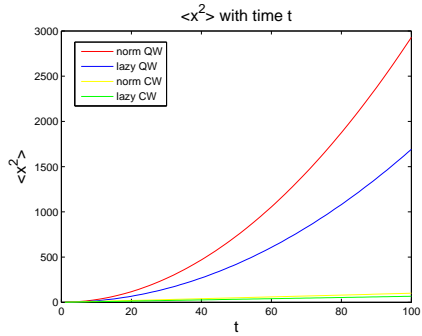}\\
    \center
  \caption{ $\langle x^2\rangle$ with time $t$. The initial states for quantum walks are $[\sqrt{0.85};0;-\sqrt{0.15}]$ and $[\sqrt{0.85};-\sqrt{0.15}]$ respectively.}\label{F32}
\end{figure}

In Fig. \ref{F32} we plot $\langle x^2\rangle$ for the lazy quantum walk, normal quantum walk, lazy classical walk and  normal classical walk. We still use the states $[\sqrt{0.85};0;-\sqrt{0.15}]$ and $[\sqrt{0.85};-\sqrt{0.15}]$ as the initial states for the two kinds of quantum walks.

From Fig. \ref{F32} , we see that though the lazy quantum walk has lower values, the values of $\langle x^2\rangle$ are of the same order $O(t^2)$, which will be prove later. Also, both classical walks have the same order of $\langle x^2\rangle$, which is $O(t)$. Furthermore, whether quantum or not, lazy walks have lower values of $\langle x^2\rangle$ than non-lazy ones. The behavior obviously comes from the lazy action, which stops the walk from going far.

Now we will prove that the probability distributions from lazy quantum walks and normal quantum walks have moments of the same order, defined by
\begin{equation}\label{00}
    \langle x^n\rangle=\sum_{x} x^n P(x),
\end{equation}
where $P(x)$ is the probability of finding the particle at position $x\in \mathcal{Z}$.

\bigskip
\textbf{Theorem 1}
The $n$-th moment of the probability distribution of a $t$-step Lazy quantum walk behaves like $O(t^n)$.

\textit{Proof}
Firstly,
{\setlength\arraycolsep{2pt}
\begin{eqnarray}
\langle x^n \rangle & = & \sum_x x^n \times P(x)=\sum_x x^n {\Psi}^{\dagger}(x,t)\Psi(x,t)
    \nonumber\\
    & = & \sum_x x^n \int \frac{dk}{2\pi} \widehat{\Psi}^\dagger(k,t) e^{ikx} \int \frac{dk'}{2\pi} \widehat{\Psi}(k',t) e^{-ik'x}
    \nonumber\\
    & = &  \int \frac{dk\cdot dk'}{(2\pi)^2} \sum_x x^n e^{i(k-k')x} \widehat{\Psi}^\dagger(k,t)  \widehat{\Psi}(k',t).
\hspace{1mm}   \label{37}
\end{eqnarray}}Using the relation
{\setlength\arraycolsep{2pt}
\begin{eqnarray}
\frac{1}{2\pi} \sum_{y} y^m\cdot e^{-i(j'-j)y}=(-i)^m \delta^{(m)}(j'-j),
\hspace{1mm}   \label{38}
\end{eqnarray}}where $\delta(\cdot)$ is the Dirac function, we obtain
{\setlength\arraycolsep{2pt}
\begin{eqnarray}
\langle x^n \rangle & = & \int\frac{dk}{2\pi}(-i)^n \widehat{\Psi}^\dagger(k,t)\int \delta^{(n)}(k'-k) \widehat{\Psi}(k',t) dk'.
\hspace{1mm}   \label{39}
\end{eqnarray}}
Since
{\setlength\arraycolsep{2pt}
\begin{eqnarray}
\int f(x)\frac{d^n}{dx^n}\delta(x-c) dx=(-1)^n \left.\left[\frac{d^n}{dx^n} f(x)\right] \right|_{x=c},
\hspace{1mm}   \label{390}
\end{eqnarray}}
{\setlength\arraycolsep{2pt}
\begin{eqnarray}
\langle x^n \rangle & = & \int_{-\pi}^{\pi} \frac{dk}{2\pi} \widehat{\Psi^\dagger}(k,t) (i\frac{d}{dk})^n \widehat{\Psi}(k,t).
\hspace{1mm}   \label{391}
\end{eqnarray}}

We now insert equation \ref{36} into equation \ref{391}. Due to unitarity of $U$, the eigenvalues of $\widehat{U}$ are $\lambda_j=e^{i\omega_j(k)}$, with corresponding eigenvectors $|v_j\rangle$, $j\in \{0,1,2\}$. Therefore,
{\setlength\arraycolsep{2pt}
\begin{eqnarray}
\widehat{\Psi}(k,t)=\sum_{j}e^{i\omega_jt} \langle v_j|\widehat{\Psi}_0\rangle |v_j\rangle,
 \hspace{1mm}   \label{392}
\end{eqnarray}}{\setlength\arraycolsep{2pt}
\begin{eqnarray}
\frac{d^n\widehat{\Psi}(k,t)}{dk^n}=\sum_{j} (it)^n \frac{d^n\omega_j}{dk^n} e^{i\omega_jt} \langle v_j|\widehat{\Psi}_0\rangle |v_j\rangle+O(t^{n-1}),
 \hspace{1mm}   \label{393}
\end{eqnarray}}{\setlength\arraycolsep{2pt}
\begin{eqnarray}
\langle x^n \rangle & = & (-t)^n \int_{-\pi}^{\pi} \frac{dk}{2\pi} \sum_{j=0}^3 |\langle v_j|\widehat{\Psi}_0\rangle|^2 \frac{d^n\omega_j}{dk^n} + O(t^{n-1}).
 \nonumber\\
 &&
 \hspace{1mm}   \label{394}
\end{eqnarray}}

By substituting the eigenvalues and the eigenvectors of $\widehat{U}$ into equation (\ref{394}), we obtain the value of each moment of the lazy quantum walk. For a quantum walk with coin operator $G(\rho)$, one of the eigenvalues is 1, which causes localization effect. For a quantum walk with DFT coin operator, all eigenvalues of $\widehat{U}$ are nonconstant functions of $k$. Most importantly, all continuous spectrum eigenvalues  $\lambda_j$' are functions of $e^{ik}$, so $\frac{d^n\lambda_j}{dk^n}\neq 0$, which means $\frac{d^n\omega_j}{dk^n}\neq 0$. Therefore, $\langle x^n \rangle=O(t^n)$. Furthermore we note that there is no essential difference in the expressions for $\langle x^n\rangle$ for lazy and normal quantum walks on the line. So the $n$-th moment of the probability distributions for  lazy quantum walks and normal quantum walks are both $O(t^n)$.

\section{Occupancy number and occupancy rate}
\label{sec:level4}

In Fig. \ref{F31}, we show the probability distribution of the lazy quantum walk and the normal quantum walk. From the figure, we see that the normal quantum walk has a very high probability of 0.1304 at position $-68$, but at most positions, the lazy quantum walk has a higher probability.  (We remind the reader further that the normal quantum walk has zero probability at half of the positions in its range.) Therefore,  normal quantum walks spread quickly but do not occupy all positions. To measure this aspect of the walks we introduce the concept of occupancy number (always keeping in mind that one number cannot exhibit all properties of a probability
distribution).

Firstly, we define the range of a quantum walk.

\textbf{Definition 1} For a random walk with $t$ steps, the \emph{range} (denoted by $N(t)$) is the number of
different positions that the walker could occupy.

In extreme situations, such as when the coin operator is the identity matrix, we note that the range does not change: In fact the range of a quantum walk is only dependent on the shift operator and
the number of steps, and is independent of the coin operator and initial state. For example, a normal quantum walk on the line with $t$ steps has range $N(t)=2t+1$. But a quantum walk whose move choice is to stay or move right has range $N(t)=t+1$.

Now, we give the definition of occupancy number and occupancy rate.

\textbf{Definition 2} For a quantum walker walking on a graph, if the walker has range $N$, we define the
occupancy number to be
\begin{equation}\label{41}
    Occ(N,t)=\#\{x|P(x,t)\geq\frac{1}{N}\}
\end{equation}
where $P(x,t)$ is the probability the walker is at position $x$ after $t$ steps.

\begin{figure}[hbt]
    \center
  \includegraphics[width=6cm]{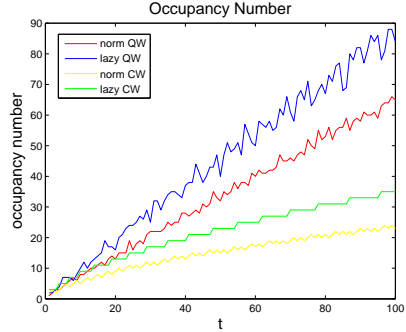}\\
    \center
  \caption{ Occupancy number with time $t$. The initial states are $[\sqrt{0.85};0;-\sqrt{0.15}]$, $[\sqrt{0.85};-\sqrt{0.15}]$ respectively for two kinds of quantum walks.}\label{F41}
\end{figure}

In the Fig. \ref{F41}, we show the dependence of the occupancy number on the number of steps $t$ (remember that the walker's  range is $2t+1$). The initial state is still $[\sqrt{0.85};0;-\sqrt{0.15}]$ and $[\sqrt{0.85};-\sqrt{0.15}]$. It is clear from the graph that the occupancy number of the quantum walk increases linearly with time $t$, though with fluctuations. This result will be proven later. In addition, the rate of increase for the lazy quantum walk is larger than that for the normal quantum walk. The rate of increase for the lazy classical walk is also larger than that for the normal classical walk. So we see that the lazy action results in a higher occupancy number.

The lazy action of the quantum walker brings some new features to the walk. The lazy walker will not have a high probability of being found at a remote position.
But from another point of view, the lazy walker is more deliberate. The walker doesn't like to move, and once he moves, he doesn't like to move again. So the walker has a higher probability (relative to the non-lazy case) of occupying positions that were previously occupied. Furthermore, quantum features (which bring coherence and decoherence)  makes the walker travel further and gives the walker $O(t^n)$ order for the moments. In summary, for a lazy quantum walk, the walker has
 \begin{itemize}
\item  $O(t^n)$ order of the $n$-th moment;
\item  a high occupancy number (relative to non-lazy case);
\item  similar probability distribution concentrated interval to that of a normal quantum walk.
\end{itemize}

Because the occupancy number of a quantum walk increases linearly with the time $t$, for ease of comparison, we define the occupancy rate as the normalized occupancy number:

\textbf{Definition 3} For a quantum walker walking on a graph, if the walker has range $N$, then we
define the occupancy rate to be
\begin{equation}\label{42}
    OccRate(N,t)=\frac{Occ(N,t)}{N}.
\end{equation}

This definition means that occupancy rate is the quotient of the occupancy number and the  range.

In fact,  the group-velocity density has the form
\begin{equation}\label{E601}
\omega(v)=\frac{\sqrt{1-\rho^2}(d_0+d_1v+d_2v^2)}{2\pi(1-v^2)\sqrt{\rho^2-v^2}}
\end{equation}
for lazy quantum walks with coin operator $G(\rho)$, while the norm quantum walks with general coin operator\begin{equation}\label{E602}
    \left(
      \begin{array}{ccc}
        a & b  \\
        c & d  \\
      \end{array}
    \right)
\end{equation} has the following density, known as Konno's density function
\begin{equation}\label{E603}
\omega(v)=\frac{\sqrt{1-|a|^2}(1+d_1v )}{\pi(1-v^2)\sqrt{|a|^2-v^2}},
\end{equation}where $d_0,d_1,d_2$ are parameters dependent on the coin operator and initial state. Because\begin{equation}\label{E604}
    lim_{t\rightarrow +\infty}\frac{m}{t}=v,
\end{equation}with the group-velocity density, by integrating the group-velocity density on the intervals satisfying $P(x,t)\geq\frac{1}{2}$, we can easily get the approximate limit of occupancy rate. Due to the result of the integration is a constant number, $OccRat$ has the order $O(1)$.

$OccRate(N,t)$ has the following properties:
\begin{itemize}
  \item  $\frac{1}{N} \leq OccRate(N,t) \leq 1$ for arbitrary $N$ and $t$;
  \item $OccRate(N,t)=1$ when the probabilities at every position are identical (and equal to $\frac{1}{N}$);
  \item $OccRate(N,t)=\frac{1}{N}$ when all probabilities satisfy $P(x,t) < \frac{1}{N}$ except at one position.
\end{itemize}

For example, after 50 steps, the occupancy number for the lazy quantum walk in Fig. \ref{F41} is 40, while the range is $2(50)+1=101$. So the occupancy rate for the quantum walk after 50 steps is $40/101\approx0.396$.
\begin{figure}[hbt]
    \center
  \includegraphics[width=6cm]{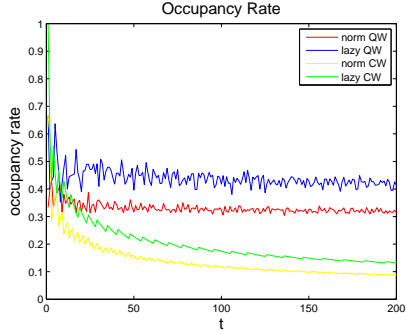}\\
    \center
  \caption{ Occupancy rate plotted against time $t$. The initial states are $[\sqrt{0.85};0;-\sqrt{0.15}]$ and $[\sqrt{0.85};-\sqrt{0.15}]$ respectively for lazy quantum walks and normal quantum walks.}\label{F42}
\end{figure}

In the Fig. \ref{F42}, we show the dependence of the occupancy rate on the time $t$. For quantum walks, since the occupancy number increases linearly with the time $t$, the occupancy rate is $O(1)$. Also, from Fig. \ref{F42}, it is clear that the occupancy rate of the quantum walk fluctuates around a fixed value. In general, lazy quantum walks have higher occupancy rate than other corresponding norm quantum  walks on the line, and this behaviour also holds for the occupancy number. This is easy to understand since, for time $t$, all occupancy rates are the quotients of the respective occupancy numbers and the quantity $(2t+1)$. So, the features of the occupancy rate are the same as those of the occupancy number, except the order becomes $O(1)$. (We should remind readers that the occupancy rate of a classical walk decreases asymptotically to $0$ (see the Appendix).)

We now pose a further question: Do quantum walks with different coin operators all have high occupancy rate?

\begin{figure}[hbt]
    \center
\includegraphics[width=5cm]{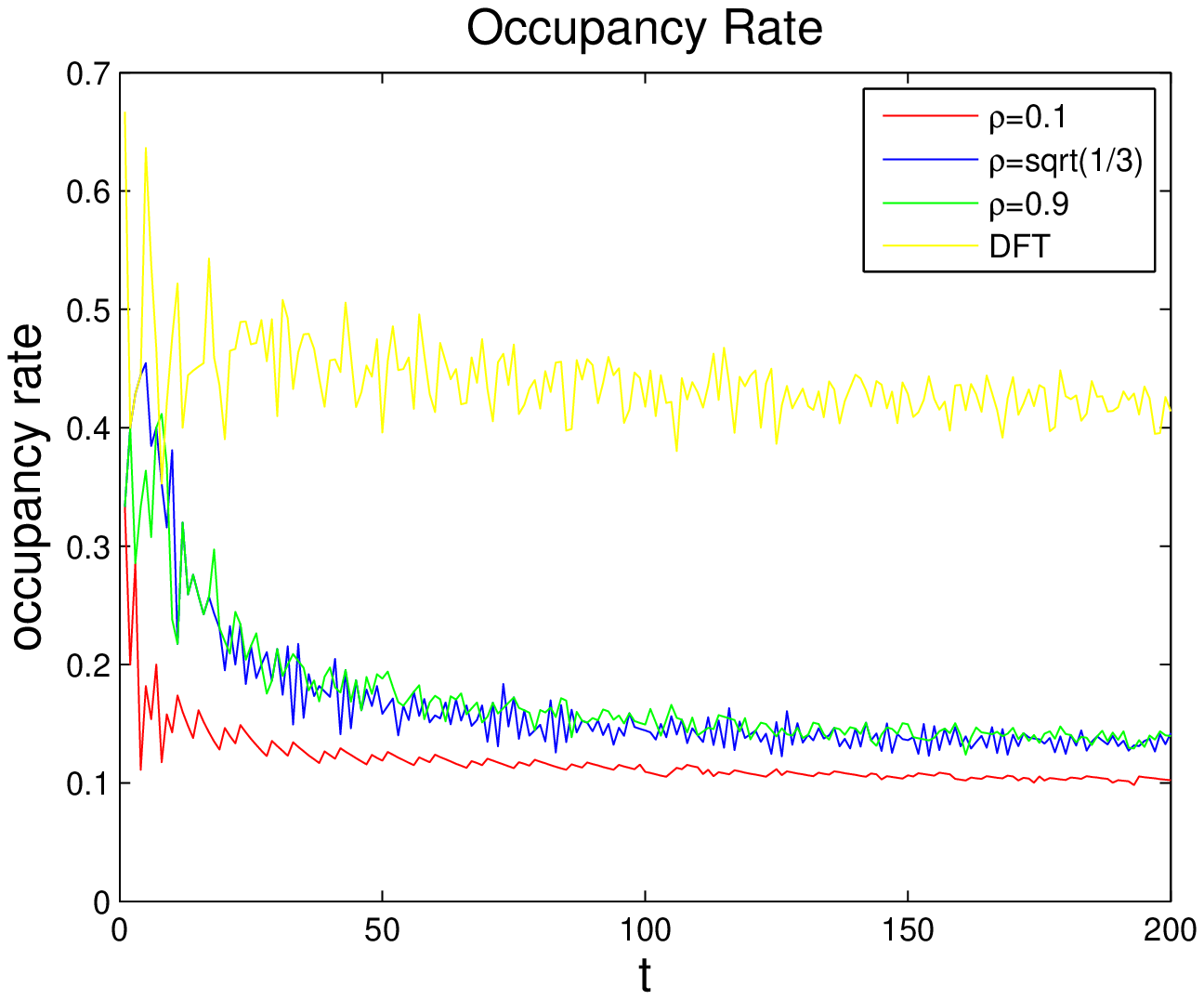}
\includegraphics[width=5cm]{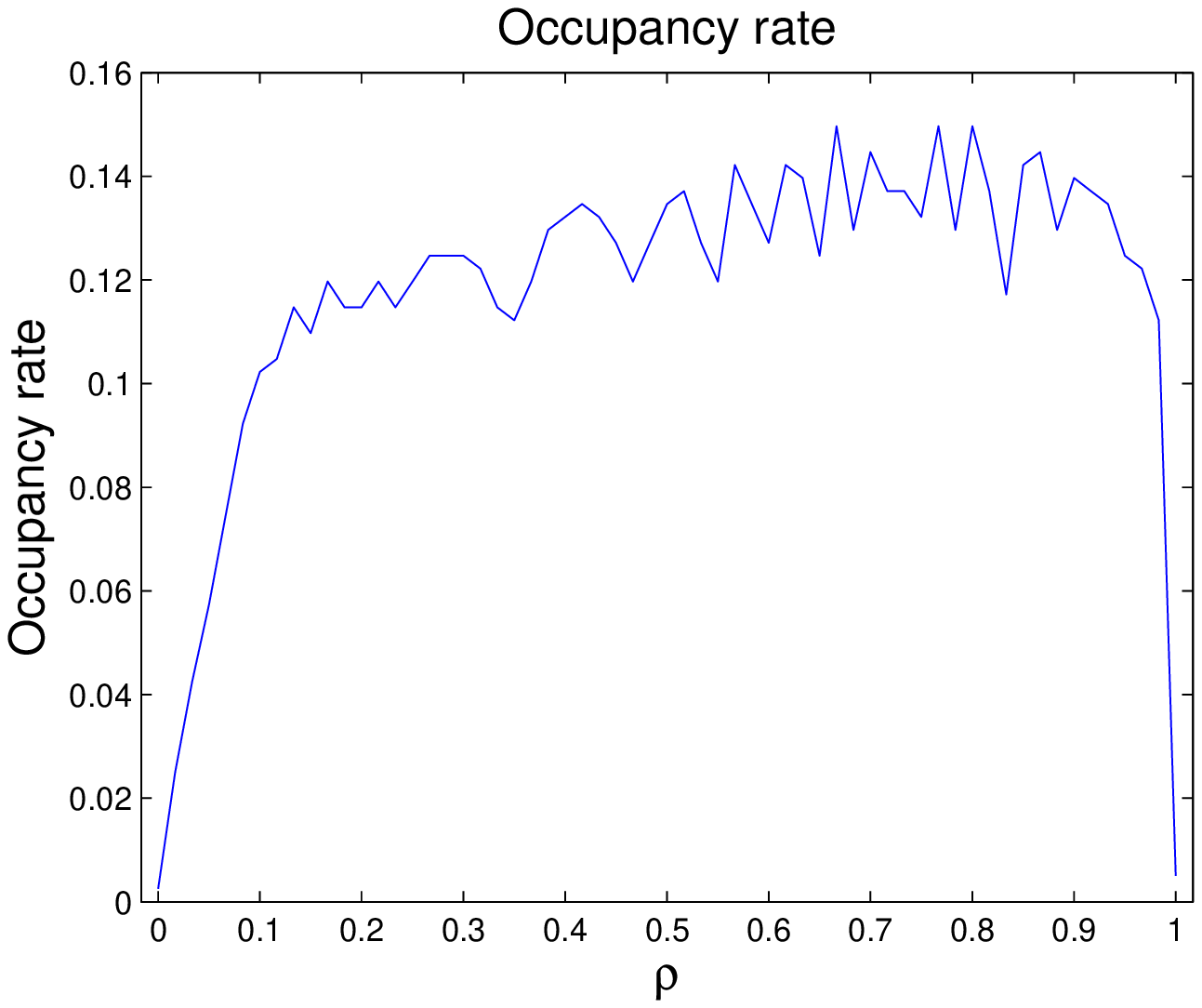}
    \center
  \caption{ Subgraph (a) shows the occupancy rate with time $t$.  The initial states are $[\sqrt{0.85};0;-\sqrt{0.15}]$, $[\sqrt{0.85};-\sqrt{0.15}]$ respectively for lazy and non-lazy quantum walks. Subgraph (b) shows the occupancy rate at time 200 with laziness $\rho$.}\label{F44}
\end{figure}

In Fig. \ref{F44}, we plot the occupancy rate versus the time $t$. In subgraph (a), we choose the matrix $G$ in equation \ref{24} (with $\rho=0.1,sqrt(1/3),0.9$) and the DFT matrix as coin operators. In subgraph (b), we show the change of occupancy rate at time 200 with laziness $\rho$.  We find that the occupancy rate will increase slowly with varying laziness parameter $\rho$ except the small region near $\rho=0 or 1$. Because we choose the occupancy rate at time 200, the occupancy rate in subgraph (b) is fluctuant. We also find that, even though the occupancy rate varies with $\rho$, the maximum value of the occupancy rate for this kind of lazy quantum walk is less than 0.2, which is lower than that for the quantum walk with DFT coin operator. We also note the quantum walk with coin operator $G(\rho=-sqrt(\frac{1}{3}))$ has lower occupancy rate than that for the hadamard quantum walk. Does this mean that higher occupancy rate is not a general property of lazy quantum walks? We remind readers that the coin operator $G$ comes from Grover coin operator, and the corresponding $2\times2$ Grover matrix is the Pauli $X$ matrix, whose occupancy rate is close to 0.

In more general scenarios, we may want to parameterize the threshold probability beyond which
we say a position is ``occupied''.  So, borrowing ideas from fuzzy set theory, we define the general occupancy number and general occupancy rate using a parameter $\delta$ which satisfies $0 < \delta \leq N$.

\textbf{Definition 4} For a quantum walker walking in one dimension, if the walker has range $N$, then we define the \emph{general occupancy number} to be
\begin{equation}\label{42}
    GenOcc(\delta,N,t)=\#\{x|P(x,t)\geq\frac{\delta}{N}\}.
\end{equation}

\textbf{Definition 5} For a quantum walker walking in one dimension, if the walker has range $N$, then we define the \emph{general occupancy rate} to be
\begin{equation}\label{42}
    GenOccRate(\delta,N,t)=\frac{GenOcc(\delta,N,t)}{N}.
\end{equation}

The general occupancy rate has the following properties:
\begin{itemize}
  \item  $0 \leq GenOccRate(\delta,N,t) \leq 1$ for all $\delta,N,t$;
  \item If $\delta1<\delta2$, $GenOccRate(\delta1,N,t) \geq GenOccRate(\delta2,N,t)$.
\end{itemize}

\begin{figure}[hbt]
    \center
  \includegraphics[width=6cm]{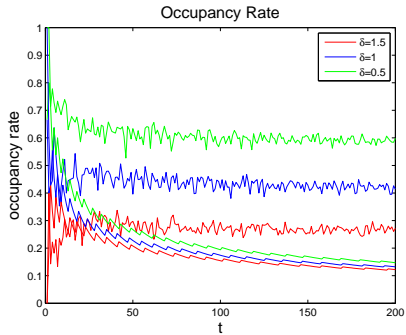}\\
    \center
  \caption{ Dependence of the general occupancy rate on the time $t$ for a 200-step walk. Red lines represent the lazy quantum walk and lazy classical walk with $\delta=1.5$. Blue lines represent the lazy quantum walk and lazy classical walk with $\delta=1$. Green lines represent the lazy quantum walk and lazy classical walk with $\delta=0.5$. }\label{F45}
\end{figure}

In Fig. \ref{F45}, we show the general occupancy rate of quantum walks and classical walks for
different values of the parameter $\delta$. Blue lines represent the general occupancy rates for quantum walks and classical walks with $\delta=1$, i.e the occupancy rates previously considered in Fig. \ref{F42}. From Fig. \ref{F45}, we see that  properties of the general occupancy rate don't change with the parameter $\delta$. For quantum walks, the general occupancy rate still converges, with small fluctuations, to a non-zero value, while for classical walks it decreases asymptotically to 0. In addition, lazy classical walks and normal classical walks have identical  orders for their occupancy rates, and the general occupancy rate for both quantum walks and classical walks is $O(t)$. We show (in the Appendix) that for normal classical walks, the occupancy rate is $O(\sqrt{t})$. Therefore, the general occupancy rate for classical walks is
also $O(\sqrt{t})$. Furthermore, the general occupancy rate is larger if $\delta$ smaller, for all kinds of walks.

\section{Entanglement between position and coin}
\label{sec:level5}

Entanglement is believed to be the most important quantum resource, occurring only in quantum states. For a single particle quantum walk, the entanglement between position and coin is not negligible \cite{501,502}. In this section, we study the entanglement between  position and coin for lazy quantum walks on the line. We use the standard entanglement measure (von Neumann entropy) to measure the total entanglement between the two subsystems, position space and coin space.

The entanglement between two subsystems of a bipartite pure quantum state $|\Psi\rangle$ can be quantified using the von Neumann entropy $S$ of the reduced density matrix of either subsystem,
{\setlength\arraycolsep{2pt}
\begin{eqnarray}
E(|\Psi\rangle)=S(\rho_1)=S(\rho_2)=-Tr(\rho_1log_2\rho_1) = -Tr(\rho_2log_2\rho_2),
\hspace{1mm}   \label{51}
\end{eqnarray}}
where $\rho_1, \rho_2$ are, respectively, the reduced density matrices of systems 1, 2, and
 $0log_20=0$.

\begin{figure}[hbt]
    \center
  \includegraphics[width=6cm]{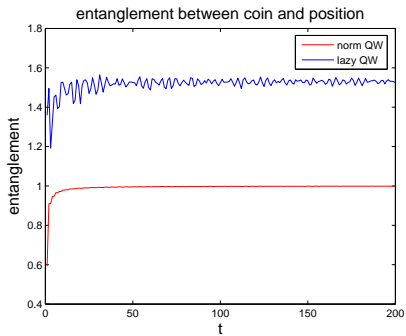}\\
    \center
  \caption{Dependence of  $\langle x^2\rangle$ on time $t$. The initial states are $[\sqrt{0.85};0;-\sqrt{0.15}]$ and  $[\sqrt{0.85};-\sqrt{0.15}]$  for lazy and normal quantum walks, respectively.}\label{F51}
\end{figure}

Numerical methods can then be used to calculate the entanglement $E$ between position and coin. In Fig. \ref{F51}, we show the entanglement for normal quantum walks and lazy quantum walks. The maximum value of the entanglement $E$ between two subsystems is $E_{max}=log_2(min(d_1,d_2))$, where $d_1$ and $d_2$ are the dimensions of the two subsystems. For normal single-particle quantum walks on the line, $E_{max}=log_2(2)=1$, while for lazy single-particle quantum walks on the line $E_{max}=log_2(3)\approx 1.585$. From Fig. \ref{F51}, we see that the entanglement converges to the maximum value for the two kinds of quantum walks. Because the dimension of the coin space for lazy quantum walks is bigger than that for normal quantum walks, lazy quantum walks have a higher entanglement between  position and coin. Though higher entanglement is a quantum resource, this results from the ``cost'' of constructing a 3-dimensional coin: If we want higher entanglement, we have to pay for it!

\section{Conclusion}
\label{sec:level6}

In this paper, we study properties of discrete lazy quantum walks. We discuss the probability distribution concentrated interval and the moments of lazy quantum walks. We prove that lazy quantum walks and normal quantum walks have the some order for the moments of their probability distributions, $O(t^n)$.

We introduce the  occupancy number and occupancy rate to measure the extent to which the walk has a (relatively) high probability at every position in its range. From our research, we conclude that lazy quantum walks have higher occupancy rate than other walks such as normal quantum walks, classical walks and lazy classical walks. We show that DFT lazy quantum walks and hadamard quantum walks have similar probability distribution concentrated intervals but dissimilar occupancy rates. We also discuss other coin operators. Among the coin operators we consider, quantum walks with DFT coin operator have the highest occupancy rate.

Finally, we study the entanglement between position and coin for lazy quantum walks. The entanglement for lazy quantum walks is higher than that for normal quantum walks, which is a benefit arising from the higher dimensional coin.

Based on the these properties of lazy quantum walks, we hope in future to work to find applications.

\section*{Appendix: Asymptotics of the occupancy rate for classical random walks}

For the Classical Random Walk (CRW), we have that the probability of being at position x after $t$ steps for $x\leq t$ is
\begin{equation}
P(x,t) =
\left \{
\begin{array}{cl}
  \frac{t!}{((t+x)/2)!((t-x)/2)!2^t}
 &\mbox{ if $x+t$ is even} \\
  0 &\mbox{ otherwise}
       \end{array} \right.
\end{equation}
For large $t$ and for $x << t$, the standard treatment uses the Stirling approximation
\begin{equation}
t! \approx
(t/e)^t\sqrt{2\pi t}
\end{equation}
and the Taylor Series approximation
\begin{equation}
\ln(1\pm x/t) \approx
\pm x/t - x^2/(2t^2)
\end{equation}
to give that
\begin{equation}
P(x,t) \approx
\sqrt{\frac{2}{\pi t}}e^{-x^2/2t}.
\end{equation}

Since the walk is symmetric (i.e. $P(x,t) = P(-x,t)$, the probability  for positive $x$
equals that for negative $x$), it suffices to examine only $x>0$. Since $P(x,t)$ is a strictly decreasing
function of $x$, there is some point $0 < x^* < t$ for which
\begin{equation}
P(x,t)
\left \{
\begin{array}{cl}
 < 1/(2t+1)
 &\mbox{ if $x \geq x^*$ } \\
 \geq 1/(2t+1)  &\mbox{ if $x < x^*$}
       \end{array} \right.
\end{equation}
where $1/(2t+1)$ is the average probability for a $t$ step walk.
We define the occupancy number by
\begin{equation}
Occ(t) = \#\{x | P(x,t) \geq 1/(2t+1)\}\label{occ1}
\end{equation}
and the occupancy rate by
\begin{equation}
OccRate(t) = Occ(t)/(2t+1)
\end{equation}

To determine $OccRate(t)$ we must solve $P(x^*,t) = 1/(2t+1)$ for $x^*$:
\begin{eqnarray}
  &\sqrt{\frac{2}{\pi t}}e^{-(x^*)^2/2t} = \frac{1}{2t+1}\hspace{0.5cm}
  \nonumber \\
\Rightarrow \hspace{0.5cm}
&\frac{(x^*)^2}{2t} = - \ln \left( \frac{1}{2t+1} \sqrt{\frac{\pi t}{2}}\right)
\approx - \ln \sqrt{\frac{\pi }{8t}}
 \nonumber \\
\Rightarrow \hspace{0.5cm}
&(x^*)^2 \approx -t\ln\frac{\pi}{8t}
 \nonumber \\
\Rightarrow \hspace{0.5cm}
&x^*\approx \sqrt{t}\sqrt{-\ln \frac{\pi}{8t}}
\end{eqnarray}
so that the occupancy rate is
\begin{equation}
OccRate(t) \approx \frac{x^*}{2t+1}
	= \frac{ \sqrt{t}\sqrt{-\ln \frac{\pi}{8t}}}{2t+1}
\approx
\sqrt{\frac{-\ln \frac{\pi}{8t}}{4t}}.
\end{equation}

From L'H\^opital's Rule we obtain
\begin{equation}
\lim_{t\to\infty} OccRate(t) \approx
\frac{\frac{1}{2}      (-\ln\frac{\pi}{8t})^{-1/2}/t} {t^{-1/2}}
=
\frac{1}{\sqrt{-t\ln \frac{\pi}{8t}}}
=0
\end{equation}

Note that, were we to define the occupancy number as
\begin{equation}
Occ(t) = \#\{x | P(x,t) \geq K\}
\end{equation}
instead of (\ref{occ1}), for any constant $K$, we would similarly obtain that
\begin{equation}
\lim_{t\to\infty} OccRate(t) =0
\end{equation}

\hspace{2cm}

\section*{Acknowledgments}
We acknowledge the help of Marcelo Forets about finding general $3\times3$ coin opertor. This work is supported by NSFC (Grant Nos. 61300181, 61272057, 61202434, 61170270, 61100203, 61121061), Beijing Natural Science Foundation (Grant No. 4122054), Beijing Higher Education Young Elite Teacher Project, BUPT Excellent Ph.D. Students Foundation(Grant Nos. CX201325, CX201326) and the China Scholarship Council (Grant Nos. 201306470046).

\vspace*{2mm}

%%%% 参考文献排版格式：

\end{CJK*}  %% 结束中文、日文、韩文使用环境

\begin{thebibliography}{99}
\itemsep=-4pt plus.2pt minus.2pt  %% 调整参考文献条与条之间的间距
\small
%\bibitem{1}  %% 输入参考文献1内容
%\bibitem{2}  %% 输入参考文献2内容
%\bibitem{3} %% 输入参考文献3内容，其余依次往下排

\bibitem{000} Venegas-Andraca, S E 2012  {\it Quant. Inf. Proc.} {\bf 11} 5,  pp 1015-1106

\bibitem{001} Reitzner D, Nagaj D, and Buzek V 2013 {\it arXiv:} quant-ph/1207.7283v2

\bibitem{110} Ambainis A 2003 {\it arXiv:} quant-ph/0311001

\bibitem{111} Shenvi N, Kempe J, and BirgittaWhaley K 2003 {\it Phys. Rev. A}  {\bf 67} 052307

\bibitem{112} Hein B, and Tanner G 2010 {\it Phys. Rev. A}  {\bf 82} 012326

\bibitem{113} Berry S D, and Wang J B 2010 {\it Phys. Rev. A} {\bf 82} 042333

\bibitem{114} Tarrataca L,  and Wichert A 2013 {\it Quant. Inf. Proc.} {\bf 12}  2, pp 1365-1378

\bibitem{115} Li D, Zhang J, Guo  F Z, Huang W, Wen Q Y, and Chen H 2013 {\it Quant. Inf. Proc.}  {\bf 12} 3, pp 1501-1513

\bibitem{116} Li D, Zhang J, Ma X W, Zhang W W, and Wen Q Y 2013 {\it Quant. Inf. Proc.} {\bf 12} 6, pp 2167-2176

\bibitem{117} Berry S D, and Wang  J B 2011 {\it Phys. Rev. A} {\bf 83} 042317

\bibitem{118} Douglas B L, and Wang J B 2008 {\it J. Phys. A}  {\bf 41} 075303

\bibitem{120} Ambainis A, Bach E, Nayak A, Vishwanath A, and Watrous  J 2011  {\it STOC '01 Proceedings of the thirty-third annual ACM symposium on Theory of computing} (ACM New York, NY, USA) pp. 37-49

\bibitem{121} Nayak A, and Vishwanath A 2000  {\it arXiv:} quant-ph/0010117

\bibitem{128} Chou C I, and Ho C L 2014 {\it Chin. Phys. B} {\bf 23} 110302

\bibitem{503} Li M, Zhang Y S, and Guo G C 2013 {\it Chin. Phys. B} {\bf 22} 030310

\bibitem{122} Xue P, and Sanders B C 2012 {\it Phys. Rev. A} {\bf 85} 022307

\bibitem{123} DiFranco C, McGettrick M, and Busch T 2011 {\it Phys. Rev. L}  {\bf 106} 080502

\bibitem{124} DiFranco C, McGettrick M, Machida T, and Busch T 2011  {\it Phys. Rev. A}  {\bf 84} 042337

\bibitem{002} Inui N, Konno N, and Segawa E 2005 {\it arXiv:} quant-ph/0507207v1

\bibitem{125} Rohde P P, Schreiber A, Stefanak M, Jex I, and Silberhorn C 2011 {\it New J. Phys.} {\bf 13} 013001

\bibitem{126} Mayer K, Tichy M C, Mintert F, Konrad T, and Buchleitner A 2011 {\it Phys. Rev. A} {\bf 83} 062307

\bibitem{129} Zhang R, Qin H, Tang B, and Xue P 2013 {\it Chin. Phys. B} {\bf 22} 110312

\bibitem{127} Zhang R, Xu Y Q, and Xue P 2015 {\it Chin. Phys. B} {\bf 24} 010303


\bibitem{130} Childs A M 2010 {\it Commun. Math. Phys.} {\bf 294} 581-603

\bibitem{301}  Stefanak M, Bezdekova I, and Jex I 2014 {\it arXiv:} 1405.7146v2

\bibitem{302} Falkner S, and Boettcher S 2014  {\it arXiv:} 1404.1330v2

\bibitem{304} Stefanak M, Bezdekova I, and Jex I 2012  {\it  Eur. Phys. J. D} {\bf 66} 142

\bibitem{305} Stefanak M, Bezdekova I, Jex I, and Barnett S M 2014 {\it arXiv} 1309.7835v2

\bibitem{306} Machida T 2014  {\it arXiv} 1404.1522v1

\bibitem{201} Carneiro I, Loo M, Xu X, Girerd M, Kendon V, Knight P L 2005 {\it New J. Phys.} {\bf 7} pp 156

\bibitem{003} Tregenna B, Flanagan W, Maile R, and Kendon V 2003  {\it New J. Phys.}  {\bf 5} 83

\bibitem{501} Maloyer O, Kendon V 2007  {\it New J. Phys.}  {\bf 9} 87


\bibitem{502} Annabestani M, Abolhasani M R, Abal G 2010  {\it J. Phys. A} {\bf 43} 075301




\end{thebibliography}
\end{document}